# Magnonic Interferometric Switch for Multi-Valued Logic Circuits

M. Balynsky[1], A. Kozhevnikov[2], Y. Khivintsev[2,3], T. Bhowmick[1], D. Gutierrez[1], H. Chiang [1], G. Dudko[2], Y. Filimonov[2,3], G. Liu[1], C. Jiang[1], A.A. Balandin[1] , R. Lake[1] and A. Khitun[1]

[1]Department of Electrical and Computer Engineering, University of California - Riverside, Riverside, California, USA 92521

[2]Kotelnikov Institute of Radioengineering and Electronics of the Russian Academy of Sciences, Saratov, Russia 410019

[3]Saratov State University, Saratov, Russia 410012

**Abstract:** We investigated a possible use of the magnonic interferometric switches in multi-valued logic circuits. The switch is a three-terminal device consisting of two spin channels where input, control, and output signals are spin waves. Signal modulation is achieved via the interference between the source and gate spin waves. We report experimental data on a micrometer scale prototype based on $Y_3Fe_2(FeO_4)_3$ structure. The output characteristics are measured at different angles of the bias magnetic field. The On/Off ratio of the prototype exceeds 36 dB at room temperature. Experimental data is complemented by the theoretical analysis and the results of micro magnetic simulations showing spin wave propagation in a micrometer size magnetic junction. We also present the results of numerical modeling illustrating the operation of a nanometer-size switch consisting of just 20 spins in the source-drain channel. The utilization of spin wave interference as a switching mechanism makes it possible to build nanometer-scale logic gates, and minimize energy per operation, which is limited only by the noise margin. The utilization of phase in addition to amplitude for information encoding offers an innovative route towards multi-state logic circuits. We describe possible implementation of the three-value logic circuits based on the magnonic interferometric switches. The advantages and shortcomings inherent in interferometric switches are also discussed.

## I. Introduction

Spintronics has been recognized as a new emerging approach towards novel computing devices, which takes advantages of spin in addition to the electric charge [1]. The electron wave analog of the electro-optic modulator proposed by Datta and Das is one of the best known examples [2]. The schematics of the modulator and its output characteristics are shown in Fig.1(A). This is a three-terminal electronic device with the source, drain, and gate contacts, where the source-to-drain current is controlled by the application of the gate voltage. On the first look, this device is very similar to an ordinary semiconductor field effect transistor (FET). However, its principle of operation is completely different. The source and the drain are made of a ferromagnetic (or half-metallic) material working as polarizer and analyzer for the spin polarized current. The electrons are injected into a quasi-one-dimensional semiconductor channel from the magnetic source. The probability that the electrons will be transmitted through the channel/drain interface depends on the relative orientation of the electron's spin with the drain magnetization. The precession of the injected spins in the channel is controlled with a gate potential via the Rashba spin-orbit coupling effect. Thus, the source-drain current oscillates as a function of the gate voltage similar to the intensity modulation in the electro-optic modulators [2]. This work [2] has stimulated a great deal of research in the field of spintronics. The first working spin-FET prototype based on InAs heterostructure was demonstrated in 2009 [3]. This device shows an oscillatory conductance as a function of the applied voltage as was originally predicted by Datta and Das [2].

Efficient spin injection and long spin diffusion length in the channel are the two major problems inherent to all spin-FETs. Being injected into a semiconductor channel, the spins of conduction electrons are subject to different relaxation mechanisms (e.g. Elliott–Yafet [4], D'yakonov–Perel' [5], Bir–Aronov–Pikus [6]), which reduce the spin polarization. All scattering mechanisms tend to equalize the number of spin up and spin down electrons in a non-magnetic semiconductor channel. In turn, the variation of the spin polarization among the ensemble of conducting electrons reduces the On/Off ratio. In the best scenario, materials with the high mobility and low scattering (e.g. graphene) show electron spin diffusion length of the order of several micrometers at room temperature [7]. The problems associated with limited spin diffusion length can be resolved by utilizing collective spin phenomena, where the interaction among a large number of spins makes the system more immune to the scattering. A spin wave is a collective oscillation of spins in a lattice around the direction of magnetization. Spin waves appear in magnetically ordered structures, and a quantum of spin wave is referred to as a magnon. The collective nature of spin wave phenomena manifests itself in relatively long coherence length, which may be order of the tens of micrometers in conducting ferromagnetic materials (e.g. $Ni_{81}Fe_{19}$ [8]) and exceed millimeters in non-conducting ferrites (e.g. $Y_3Fe_2(FeO_4)_3$ [9]) at room temperature. The first working spin wave-based logic device was experimentally demonstrated in 2005 by Kostylev et al. [10]. In this work, the authors built a Mach–Zehnder-type spin wave interferometer to demonstrate the output voltage modulation as a result of spin wave interference. The schematics and the output characteristics of the spin wave device are shown in Fig.1(B). The phase difference among the spin waves propagating in the arms of the interferometer is controlled by the magnetic field produced by the electric current $I_G$. At some point, the output characteristics of this device resemble the ones of the Datta and Das device, while the oscillation of the output voltage is controlled by the magnetic field (gate current).

Later on, exclusive-not-OR and not-AND gates have been experimentally demonstrated on a similar Mach-Zehnder-type structure [11]. However, the integration of several of such devices in a circuit requires spin wave-to-magnetic field conversion, which dictates a need in additional converter circuits. This issue has been addressed in the magnon transistor demonstrated by Chumak et al in 2013 [12]. The schematics and the output characteristics of the magnon transistor are shown in Fig. 1(C). The transistor is based on a magnonic crystal designed in the form of an yttrium iron garnet (YIG) film with an array of parallel grooves at its surface. The magnons are injected into the transistor's source and are detected at the drain using microstrip antennas. The magnons that control the source-to-drain magnon current are injected directly into the magnonic crystal (transistor's gate). The principle of operation is based on the nonlinear four-magnon scattering mechanism, which makes it possible to attenuate source-drain transport when gate magnons are injected. The injection of the gate magnons suppresses the source-drain magnon current as illustrated in Fig.1(C).

In this work, we describe a magnonic switch based on the spin wave interference. We argue that the relatively simple interference-based device possesses output characteristics similar to the Datta and Das device Fig.1(A). The proposed device makes it possible to exploit both the phase and amplitude of the output as in the spin wave Mach-Zehnder interferometer shown in Fig.1(B), and allows one to build all-magnon logic circuits similar to the magnon transistor shown in Fig.1(C). The rest of the paper is organized as follows. In Section II, we describe the material structure and the principle of operation of a magnonic switch based on the interference effect. We describe possible multi-valued logic circuits based on the magnonic switch. In Section III, we present experimental data obtained for a micrometer scale prototype based on YIG structure. Theoretical analysis and results of micromagnetic simulations are given in Section IV. We also present the results of numerical simulations illustrating the operation of the ultimately scaled switch consisting of two 1-D spin channels. The discussion and conclusions are given in Sections V and VI, respectively.

## II. Principle of Operation

The schematic of the magnonic interferometric switch is shown in Fig.2(A). It is a three-terminal junction comprising the two lines of spin channels. To be consistent with the examples of spin-based devices shown in Fig.1, we depict the three terminals as a "source", a "gate", and a "drain". The source-drain channel is a chain of spins coupled via the exchange interaction. The gate is connected to the source-drain channel via the similar one-dimensional chain of spins. These two chains are shown orthogonal to each other in Fig.2(A) though the angle of the junction is not of critical importance. The input signal coming from the source is a continuous spin wave $m_S(r,t)$ of fixed frequency $f$, amplitude $A_S$, and phase $\varphi_S$:

$$m_S(r,t) = A_S \cdot exp[-\kappa r] \cdot sin(k_0 r - \omega t + \varphi_S) \tag{1}$$

where $\kappa$ is the damping constant, $r$ is the distance traveled, $k_0$ is the wave vector, $\omega = 2\pi f$, and $t$ is the time.

The control signal is also a continuous spin wave of the same frequency $f$, amplitude $A_G$, and phase $\varphi_G$. These two waves propagate through the spin chains and reach the drain. The output of the device is a spin wave – a result of the spin wave interference:

$$m_D(r,t) = m_S(r,t) + m_G(r,t) \tag{2}$$

The amplitude $A_D$ and phase $\varphi_D$ of the output depend on the phase difference $\Delta\varphi_{SG} = (\varphi_S - \varphi_G)$. The output has maximum amplitude if the spin waves are coming in phase (constructive interference = On state). The output has minimum amplitude if the waves are coming out of phase (destructive interference = Off state). In Fig.2(B), we depicted the amplitude of the output $A_D$ as a function of the source-gate phase difference $\Delta\varphi_{SG}$ assuming the waves reach to the drain with the same amplitude (e.g. $A_S = A_G$).

Conventional digital logic circuits operate with a two-valued logic where 0 and 1 correspond to the two levels of voltage or current (i.e. On and Off states of FET). The output characteristics of the magnonic interferometric switch is also suitable for amplitude encoding, where On and Off states correspond to the cases of constructive and destructive interference. In theory, the On/Off ratio of the magnonic switch may be infinity, as there is zero amplitude output in the case of the destructive interference. In addition to the amplitude, we can also utilize the phase of the output as an additional logic variable. For example, there may be several On states with the same amplitude but different phase $\varphi_D$ (e.g. 0 or $\pi$). The benefit of having phase for logic encoding can be utilized for multi-valued logic [13].

In order to illustrate this idea, we choose five output states as depicted in Fig.2(C) : [-2 logic state: $A_D = 2A_0$, $\varphi_D = 0\pi$]; [-1 logic state: $A_D = A_0$, $\varphi_D = 0\pi$]; [0 logic state $A_D = 0$, $\varphi_D$ is not defined]; [+1 logic state: $A_D = A_0$, $\varphi_D = \pi$]; [+2 logic state: $A_D = 2A_0$, $\varphi_D = \pi$]. Sign plus or minus corresponds to the phase of the output (e.g. 0 or $\pi$), while the digit correspond to the amplitude of the output: $2A_0$, $1A_0$, or $0A_0$, where $A_0$ is some reference amplitude). The reason of taking these particular states lies in the specific of interferometric switching. Let us consider the operation of magnonic switch as shown in Fig.2(A), in case of only three possible states: -1,0 and +1 for the source and the gate. There are nine possible input state combinations resulting in five possible output states (e.g. -2,-1,0,+1,+2) as summarized in the Table in Fig.2(c). The applying of the 5-states input to the interferometric switch will provide output with 9 possible states (e.g. two possible phases and five possible amplitudes). In this work, we consider an example of the three-valued (3VL) logic gates [14]. There are two more elements required for 3VL logic gate construction, which are a π-phase shifter; and an attenuator as shown in Fig.4(A). These are two passive non-linear elements for independent phase and amplitude modulation. The π-phase shifter provides a π-phase shift to the propagating spin waves. It may be an additional chain of spins (i.e. similar to the delay lines used in optics [15]), or a permanent magnet placed near the chain, or a resonator [16]. The combination of the magnonic interferometric switch and the phase shifter provides the NOT gate (i.e. the truth table in Fig. (3A)). A π-phase shift is equivalent to the 3VL Inverter logic operation -1→+1; 0→0; +1→-1. The attenuator is also a passive device (e.g. similar to a non-linear resistor in electric circuits), which reduces the amplitude of the transmitted spin wave signal 2A→A, A→0, 0→0. The introduction of the attenuator reduces the number of possible output states from 5 to 3. The combination of the attenuator with the magnonic switch makes it possible to realize 3VL XOR gate as depicted in Fig.3(B). The combination of the NOT and the XOR gates allows us to build all other types of logic gates similar to the conventional two-valued Boolean logic [17].

### III. Experimental data

In this Section, we present experimental data obtained on the available four-terminal cross junction made of yttrium iron garnet $Y_3Fe_2(FeO_4)_3$ . The photo of the device and connection schematics is shown in Figure 4. The cross junction is made of single crystal YIG film epitaxially grown on top of a Gadolinium Gallium Garnett ($Gd_3Ga_5O_{12}$) substrate using the liquid-phase transition process. After the films were grown, micro-patterning was performed by laser ablation using a pulsed infrared laser (λ≈1.03 μm), with a pulse duration of ~256 ns. The YIG cross has the following dimension: the length of the each waveguide is 3.65 mm; the width is 650 μm; and the YIG film thickness is 3.8 µm; and saturation magnetization of $4\pi M_0 \approx 1750\, Oe$. There are four Π-shaped micro-antennas fabricated on the edges of the cross. Antennas were fabricated from a gold wire of thickness 24.5µm and placed directly at the top of the YIG surface. The antennas are connected to a programmable network analyzer (PNA) Keysight N5241A. Two of the antennas depicted by the letters S and G are used to generate the source and the gate spin waves, respectively. The inductive voltage is detected by the "drain" antenna as depicted in Fig. 4. Spin waves were excited by the magnetic field generated from the AC electric current flowing through the S and G antenna(s). We used a set of attenuators (PE7087) and a phase shifter (ARRA 9428A) to independently control input power and the phase difference between the input ports. The inductive voltage produced by the interfering spin waves was detected by the drain antenna D. The details of the inductive measurement technique can be found elsewhere [18]. In our experiments, we used the attenuator and the amplifier depicted in Fig.4 to equalize the output voltages produced by the S and the G antennas operating separately.

We carried out three sets of experiments aimed to show the output inductive voltage modulation by the phase difference between the source and the gate spin waves. The experiments were accomplished at three different directions of the bias magnetic field. The direction of the bias magnetic field significantly affects the dispersion of the propagating spin waves. For instance, magnetostatic spin waves propagating parallel to the bias magnetic field (so-called backward volume magnetostatic spin waves BVMSW) possess negative group velocity $v_g = \partial\omega/\partial k < 0)$ , while spin waves propagating perpendicular to the bias magnetic field (so-called magnetostatic surface spin waves MSSW) possess positive group velocity $v_g > 0)$ [19]. It is critically important for the operation of the cross-type magnonic devices to ensure the propagation of the both types of waves. The latter is possible by finding a proper combination of the operational frequency and the bias magnetic field. Prior to the experiments, we found such a combination corresponding to the operational frequency *f* = 4.095 GHz, and bias magnetic field *H* = 798 Oe. All experiments are done at room temperature.

In Fig. 5(A), we present experimental data for the bias magnetic field *H* directed parallel to the virtual *S-D* line. In this case, there is a BVMSW type of spin wave propagating from the source to the drain, and a MSSW type of spin wave propagating from the gate to the S-D channel. The input power of the S-antenna is = -18 dBm, the power of the G-antenna is -12 dBm. The red and the blue markers show the amplitude and the phase of the output voltage as a function of the phase difference between the S and the G antennas $\Delta\varphi_{SG}$. We collected experimental data for 22 points with different $\Delta\varphi_{SG}$ in the range from 0 to $2\pi$. The accuracy of the phase detection by PNA is $0.008\pi$. The amplitude of the inductive voltage has maxima about 0.448 mV, while the minimum output inductive voltage below 0.02 mV. The accuracy of the inductive voltage measurements at minima is ± 0.00046 mV. The maximum output voltage corresponding to the constructive spin wave interference is about two times large than the inductive voltage produced by just one operating antenna (i.e. the inductive voltage produced by only S operating antenna is 0.228 mV, the inductive voltage produced by only G operating antenna is 0.237 mV). The red curve in Fig.5(a) shows the results of numerical modeling in the ideal case where S and G spin waves have zero phase/amplitude variation. The blue markers in Fig.5(a) depict the measured phase of the output signal. The phase of the output $\Delta\varphi_{SG}$ is defined with respect to the phase of the S input. There is a certain phase shift between the phases of the source and the drain spin waves. The phase difference is the sum of two:

$$\Delta\varphi_{DS} = \Delta\varphi_r + \Delta\varphi_{int}, \qquad (3)$$

where $\Delta\varphi_r$ is the phase change during the spin wave propagation (i.e. $\Delta\varphi_r = r/v_p$, $v_p$ is the phase velocity), and $\Delta\varphi_{int}$ is the phase change due to the phase interference. The first term in Eq.(3) does not depend on the input phases of the waves generated at the S and G terminals, while the second term is responsible for the output phase oscillation ± π/2. There is a good agreement between the experimental and theoretical data. The observed discrepancy is mainly due to the variation of the input amplitude and phases of the interfering spin waves.

Next, we carried out similar experiments for the bias magnetic field *H* directed in-plane but perpendicular to the virtual S-D line. In this case, there is a MSSW type of spin wave propagating from the source to the drain, and a BVMSW type of spin wave propagating from the gate to the S-D channel. The input power of the S-antenna was set to -24 dBm, and the power of the G-antenna was set to -6 dBm. As it the previous example, we measured the inductive voltage as a function of the phase difference between the source and the gate antennas. The red markers and the red curve in Fig.5(B) show experimental data and the results of numerical modeling, respectively. There are two maxima about 0.414 mV corresponding to the constructive spin wave interference (i.e. the inductive voltage produced by only S operating antenna is 0.225 mV, the inductive voltage produced by only G operating antenna is 0.219 mV). The minimum voltage corresponding to the destructive spin wave interference is about 0.01 mV. The blue markers and the blue curve in Fig.5(B) correspond to the experimentally measured and calculated phase of the output.

Then, we carried out experiment with the bias magnetic field *H* directed in-plane and at $45^0$ degrees to the virtual S-D line. Experimental data and the results of numerical modeling are presented in Fig.5(C). The input power of the S-antenna was set to -30 dBm, and the power of the G-antenna was set to -9 dBm. The red and the blue markers in Fig.5(C) show the amplitude and the phase of the output voltage. The maximum output inductive voltage is 0.227 mV, (i.e. the inductive voltage produced by only S operating antenna is 0.114 mV, the inductive voltage produced by only G operating antenna is 0.117 mV).

Finally, we carried out experiment in the configuration where the source and the gate antennas are located on the same arm of the cross, while the inductive voltage is measured at the orthogonal arm. This configuration is shown in the inset to Fig.5(D). The bias magnetic field $H$ is directed parallel to the virtual S-G line. The operational frequency was decreased to 3.442 GHz. Experimental data and the results of numerical modeling are presented in Fig.5(D). The input power of the S-antenna and the G-antenna was set to -13 dBm. The red and the blue markers in Fig.5(D) show the amplitude and the phase of the output voltage. The maximum output inductive voltage is about 0.49 mV, while the On/Off ratio is 22dB.

Experimental data presented in Fig.5 (A-D) demonstrate prominent output voltage modulation by the phase difference among the source and the drain signals $\Delta\varphi_{SG}$. This output characteristic is well explained by the spin wave interference. There is a good agreement between the experimental data (red and blue markers) and the results of numerical modeling (red and blue curves) in all four cases. The maximum voltage is observed in the case of constructive spin wave interference $\Delta\varphi_{SG} \approx 0$, where the amplitude of the output doubles compare to the output produced by just on operating antenna. The major discrepancy between the theoretical and experimental data is observed for $\Delta\varphi_{SG} = \pi$, which corresponds to the destructive interference between the source and the gate waves. The results of numerical modeling show zero output amplitude, while the experimental data show some finite amplitude. The On/Off ratios calculated from experimental data show: 27 dB in Fig.5(A), 32.1 dB in Fig. 5(B), 36.1 dB for the 45 degree geometry shown in Fig.5(C), and 22dB for the geometry with source and drain located on the same arm Fig.5(D). The discrepancy between the experiment and theoretical data may be attributed to many factors including structure imperfections, amplitude and phase variation of the input waves. Thermal noise is the major and fundamental factor limiting the On/Off ration of the interferometric switch. The dephasing and damping of spin waves cased to the magnon-phonon and other scattering processes will result in a non-zero output.

Every new proposed switch and logic circuit should demonstrate certain noise immunity and the electronic noise level sufficiently low for practical applications [20,21]. There are numerous internal and external noise sources, which can affect the circuit performance. New device designs and material systems used in their implementation require investigation of dominant noise mechanisms and methods for noise level reduction [22-24]. The reported data on noise in magnonic type devices is limited [25,26]. In this work, in order to study the effect of noise on the device functionality, we collected data for 15,000 subsequent measurements (150 points per 448 ms sweep), which show the output voltage variation at fixed input parameters. In Fig. 6, we present raw data collected for the configuration with the bias magnetic field $H$ directed perpendicular to the virtual S-D line (i.e. as shown in the Fig. 5(B)). The sampling rate is 334 Hz. The signal was represented as $\Delta V(t) = V(t) - V_m$, where $V(t)$ is the recorded signal and $V_m$ is the mean value of output voltage. The noise power spectral density of the output was obtained by performing FFT on $\Delta V(t)$. The extracted normalized noise spectral density, $S_V/V^2$, was around $10^{-11}$ 1/Hz at 1 Hz. This noise level is higher than the thermal noise floor defined by the Nyquist formula $S_V=4k_BTR$ (where $k_B$ is the Boltzmann constant, T is the absolute temperature and R is the resistance) but comparable to the noise level in conventional Si CMOS devices. The low noise spectral density $S_V/V^2 \sim 10^{-11}$ 1/Hz achieved in the prototype devices (not optimized for noise reduction) suggests a possibility of extremely low-noise designs achievable with the proposed magnonic interferometric switches in the multi-valued logic circuits.

**IV. Theoretical Analysis and Numerical Modeling**

Experimental data presented in the previous Section show significant difference in the device output characteristics depending on the direction of the bias magnetic field. Potentially, this effect may be utilized for building re-configurable magnetic-magnonic logic circuits [27]. The understanding of the mechanism of spin wave transport in cross-shape junctions is the key to the device optimization. As we mentioned before, spin waves excited at the source and the gate possess different dispersion (i.e. BVMSW and MSSW). In order to provide maximum spin wave transport through the junction, one has to find the region in the bias magnetic field – operational frequency space, where the both types of waves can propagate.

In the case of infinite and uniformly magnetized films, BVMSW and MSSW have non-overlapping frequency ranges $[\omega_H, \omega_0]$ and $[\omega_0, \omega_S]$, respectively, where $\omega_H = \gamma H$, $\omega_0 = \sqrt{\omega_H(\omega_H + \omega_m)}$ , $\omega_m = \gamma 4\pi M_0$, $\omega_S = \omega_H + \omega_m/2$, ($\omega_H < \omega_0 < \omega_s$), $\gamma$ is the gyromagnetic ratio. The experimentally observed coupling may be explained by taking into consideration the effect of the magnetic field anisotropy caused by the demagnetization fields in the cross junction. In order to estimate the width of the overlap region, we present estimates based on the formalism developed for a homogeneously magnetized ellipsoid [19]. The demagnetization field can be related to the magnetization of the sample by

$$\vec{H^m}(\vec{r}) = -\hat{N}(\vec{r})\vec{M}(\vec{r}), \tag{4}$$

where $\widehat{N}(\vec{r})$ is the tensor of demagnetization coefficients, which has a diagonal form for the main axes of homogeneously magnetized ellipsoid $\sum_i N_{ii} = 1$. Neglecting the non-uniformity of $\vec{M}(\vec{r})$, the frequency of the ferromagnetic resonance $\omega_0$ in the waveguides is given as follows [19]:

$$\omega_0 = \sqrt{[\omega_H + (N_{11} - N_{33})\omega_m] \cdot [\omega_H + (N_{22} - N_{33})\omega_m]}, \qquad (5)$$

where the external magnetic field *H* is directed in-plane with the ellipsoid. The demagnetization factors depend on the structure geometry. For instance, the experimental data presented in the previous Section are obtained for the cross structure with *L>>w>>d*, where *L* is the length, *w* is the width and *d* is the thickness of the YIG cross. In this case, one may restrict consideration by the width- $N_d$ and thickness-related $N_w$ demagnetization fields as $N_d$>>$N_w$>>$N_L$.

Then, for the case shown in Figure 5(A), the long-wavelength limit (i.e. FMR frequency) of the spin waves generated by the source can be found as follows:

$$\omega_0^S = \sqrt{[\omega_H + (1 - N_w)\omega_m] \cdot [\omega_H + N_w \omega_m]} = \sqrt{\omega_0^2 + N_w \cdot (1 - N_w)\omega_H \omega_m}. \qquad (6)$$

At the same time, the long-wavelength limit for the spin wave generated at the gate is given by:

$$\omega_0^G = \sqrt{[\omega_H + (1 - 2N_w)\omega_m] \cdot [\omega_H - N_w \omega_m]} = \sqrt{\omega_0^2 - N_w \cdot \omega_m (3\omega_H + 2N_w \omega_m)}. \qquad (7)$$

Eqs.(6-7), there reveal the difference in the effect of the demagnetization field $N_w$ on the FMR frequency $\omega_0$. The appearance of the demagnetization field may increase $\omega_0^S$ while decreasing $\omega_0^G$. The value of the demagnetization field can be estimated by using the on-line calculator [28], which gives $N_w = 0.012$ for the given YIG cross geometry. Then, we estimate the overlap frequency region $\Delta\omega_0^{S,G} = \omega_0^S - \omega_0^G$ for the spin wave generated by the S and G antennas (i.e. BVMSW and MSSW, respectively) as follows:

$$\Delta\omega_0^{S,G} \approx \frac{2N_w \omega_H \omega_m}{\omega_0} \qquad (8)$$

The width of the overlap region $\Delta\omega_0^{S,G}$ for the YIG cross with given geometry at the bias magnetic field *H*=1kOe is about 70 MHz. The FMR frequencies can be also estimated using OOMMF [29], which gives $\omega_0^S \approx 4.657\ GHz$ и $\omega_0^G \approx 4.595\ GHz$ ($\Delta\omega_0^{S,G} \approx 62\ MHz$). These estimates show the possibility of the frequency range overlap and provide an insight on the operational frequency range for a given structure and at certain bias magnetic field.

Figure 7 shows the results of micromagnetic simulations of propogating spin waves in a YIG cross. The direction of the bias magnetic field is the same as in Fig.5(B). The red arrow in each plot shows the direction of the input spin wave signal. There results show spin wave propagation for three frequencies: 4.50 GHz, 4.64 GHz, and 4.70 GHz, respectively. Two of these frequencies (i.e. 4.50 GHz and 4.70 GHz) are outside the overlap region $\omega_0^S > 2\pi f$ and $2\pi f > \omega_0^G$. Frequency 4.64 GHz is within the overlap region $\omega_0^G > 2\pi f > \omega_0^S$. As one can see from Fig.7, the amplitude of the transmitted signal is much higher for the case when the operation frequency is within the overlap region. The maximum spin wave propagation is observed at the frequency close to the FMR of an infinite YIG film $\omega_0 \approx 4.64\ GHz$. It should be also noted the character features of spatial amplitude distribution, which can be attributed to the effect of spin wave quantization as well as to the anisotropy of the spin waves propagating at an angle to the bias field [30].

Scaling down the size of the switch and decreasing the operation wavelength significantly changes spin wave transport. The dispersion of sufficiently short (i.e. shorter than 100nm) wavelength spin waves (so-called exchange spin waves) differ significantly from the one of the magnetostatic waves [31]. In order to illustrate the operation of a nanometer-scale magnonic interferometric switch, we present the results of numerical simulations. We consider the two perpendicular chains of spins as shown in Fig. 2(A). There are 20 spins in the chain connecting the source and the drain. There are 10 spins in the chain connecting gate and the source-drain channels. In our case, two chains intersect in just one point - one spin. The edge spins at the source and at the gate oscillate with the same frequency $f$ and amplitude $A_0$. The neighboring spins in the magnetic wires are coupled via exchange interaction, so the Hamiltonian of the system has the following form [32]:

$$H = -J\sum_{j\delta} S_j S_{j+\delta} - 2\mu H_0 \sum_j S_{jz}\ ,\qquad(9)$$

where $J$ is the exchange coupling constant with the dimension of energy, $S_j$ and $S_{j+\delta}$ are the electron-spin operators, $S_{jz}$ is the spin projection along the z direction, the index $\delta$ runs over nearest neighbors of spin $j$, $\mu$ is the magnetic moment, $H_0$ is the external magnetic field strength. The evolution equation for spin $j$ takes the following form:

$$\hbar \frac{d\vec{S}_j}{dt} = \vec{\mu} \times \vec{B}_j\ ,\qquad(10)$$

where $B_j$ is the effective magnetic field induction acting on spin $j$, which arises from the sum of the exchange field due to the coupling with the nearest neighbor spins. The detailed explanations on the one-dimensional chain model can be found elsewhere [32].

In Fig.8, we present the results of numerical modeling showing the amplitude and the phase of the output spin wave depending on $\Delta\varphi_{SG}$. The value of the exchange constant $J$ is 70 meV) [33]. The plot in Fig.8(a) shows the results of numerical modeling at zero temperature. In this ideal case, the amplitude of the transmitted is $2A_0$ in the case of the constructive interference, and zero in the case of the destructive interference (i.e. On/Off ratio is infinity). The plot in Fig.8(b) shows the results of similar simulations at room temperature (300K). The effect of the temperature has been included in the simulation through the fluctuation dissipation theorem [34,35] according to which temperature creates a random field in the propagation of the spin waves. The ratio $J_H/k_BT$ is almost equal to 3, so the temperature acts like a noise field in the simulation. The immediate result of thermal noise is a nonzero output in the case of destructive interference, and in turn, the reduction of the On/Off ratio. In general, the effect of the finite temperature on the output characteristics of the exchange-based device exponentially increases with the ratio $k_BT/J$. In Fig.9, we present the results of numerical modeling showing the noise strength as a function of the exchange coupling energy. The stronger is the exchange coupling the weaker is the effect of the thermal noise.

**V. Discussion**

There are some unique advantages and shortcoming inherent in magnonic interferometric switch. Scalability is the most appealing advantage provided by the utilization of spin waves as logic variable. In general, an interferometric switch can be realized on any type of waves (e.g. optical, acoustic, gravitational). In any case, the minimum feature size of the logic circuit is restricted by the operational wavelength $\lambda$ (i.e. the minimum size of the inverter shown in Fig.3 is $\lambda/2$). There are no fundamental physical limits restricting the scaling down the area of the switch to several square nanometers, while the practically achievable size is limited only by the capabilities of the manufacturing technique. In turn, the time delay $\tau$ scales proportional to the length of the structure (e.g. $\tau=L/v_g$, where L is the length of the channel and $v_g$ is the spin wave group velocity). For example, the time delay of the switch with 10nm long S-D, and G-D would be 1ps conservatively taking $v_g=10^4$m/s. Low energy consumption is another advantage of the proposed switch. The principle of operation of the magnonic interferometric switch is based on the spin wave interference, where On and Off states correspond to the constructive and destructive interference, respectively. In this approach, there is no fundamental limit on the minimum energy of the interfering waves except the thermal noise. In theory, the energy of the input/output waves can be in the range of the tens of $k_BT$, where $k_B$ is the Boltzmann constant and T is the temperature. The use of phase in addition to amplitude allows us to build sophisticated three-valued logic gates and construct logic circuits with a fewer number of elements than required for conventional amplitude-based circuits.

The lack of saturation region in the output characteristics is the major drawback inherent in interference-based devices including the Datta and Das modulator [2], Mach-Zehnder spin wave interferometer [10], Magnon transistor [5], as well as the described above spin wave interferometric switch. The integration of such devices in a large scale circuit is not feasible without the introduction of an additional non-linear device aimed to equalize inevitable variations of the output amplitude/phase. There are several possible solutions to this problem with a certain tradeoff between the circuit stability, speed, and energy consumption. For example, spin wave logic gates can be combined with nanomagnets, where the result of computation in the magnonic domain is translated into the state of magnetization [36]. On one hand, the combination of spin wave logic with magnetic memory may provide a route to non-volatile and imperfection prone circuits. On the other hand, the introduction of magnetic memory inside spin wave logic gate will result in an additional time delay for nanomagnet switching (e.g. ~1ns for conventional magnetic memory), which is orders of magnitude slower compare to the spin wave propagation time (e.g. order of ps in scaled spin wave logic gates). It is also possible to complement spin wave logic gates with parametric amplifiers providing both the amplitude and phase control of the output of each gate. In this scenario, there is no additional time delay as the parametric amplifiers can increase the amplitude of the propagating spin waves without an additional time delay. However, the additional power consumption on the parametric spin wave amplifiers may be orders of magnitude higher than the power consumption of spin wave logic gates themselves. At any rate, there is always a tradeoff between circuit speed, stability, and energy consumption.

Finally, we want to mention, that building the Boolean-type logic gates is not the only, and, perhaps, not the most promising way of utilizing spin wave phenomena for data processing. The ability of using spin wave interference opens a new horizon of constructing special-task data processing devices similar to ones developed in optics but at the nanometer scale. Magnonic holographic memory is one of the examples [37,38]. Exploiting spin wave phase in addition to amplitude also allows us to build more functional devices for pattern recognition [39] and prime factorization [40]. In turn, the combination of the conventional digital logic circuits with special type magnonic devices may pave the way to a novel hybrid digital-analog type of computing architecture.

**VI. Conclusions**

In this work, we considered a possibility of building spin wave interferometric switch for all-magnonic logic gates. The switch is a three-terminal device consisting of two spin channels where input, control, and output signals are spin waves. The operation of the switch is based on spin wave interference. The latter may potentially lead to the construction of nanometer scale and low power consuming magnetic logic devices. The operation of the scaled magnonic switch is illustrated by numerical modeling. The On and Off states of the switch correspond to the constructive and destructive interference, respectively.  The phase of the spin wave signal may be exploited in addition to amplitude as a logic state variable. We described an example of NOT and XOR three-valued logic gates. We also presented experimental data on a micrometer scale prototype based on $Y_3Fe_2(FeO_4)_3$ structure. Obtained data show prominent output signal modulation by the phase difference between the input spin waves. The maximum measured On/Off ratio is 36.1 dB. The normalized noise spectral density $S_V/V^2$ determined from the output signal on the order of $10^{-11}$ 1/Hz indicates a possibility of low-noise low-power designs for this technology. Potentially, magnonic switches may be exploited for building hybrid analog-digital type of computing devices.

**Acknowledgement**

This was supported by the Spins and Heat in Nanoscale Electronic Systems (SHINES), an Energy Frontier Research Center funded by the U.S. Department of Energy, Office of Science, Basic Energy Sciences (BES) under Award # SC0012670.

**Figure Captions**

Figure 1. Schematics and input-output characteristics of selected spin-based switches. (A) Electro-optic modulator proposed by Datta and Das. The flow of the spin polarized electron in the channel is controlled by the gate voltage via the Rashba spin-orbit coupling effect. The source-drain current oscillates as a function of the gate voltage. (B) Mach-Zendher spin wave interferometer. The output inductive voltage is controlled by the phase difference among the interfering spin waves. In turn, the phase difference is controlled by the magnetic field generated by DC electric current in the gate wire. Output inductive voltage oscillates as a function of the gate current. (C) Magnon transistor. The flow of magnons from the source to the drain is modulated by the magnons injected at the gate via the four-magnon scattering. The injection of the gate magnons suppresses the source-drain magnon current.

Figure 2. (A) Schematics of the magnonic interferometric switch. It is a three-terminal junction comprising the two lines of spin channels. The source, the gate and the drain terminals are depicted by letters S,G, and D. One of the spin chains serves as a source-drain channel, and the other chain connects the gate with the source-drain channel. The source and the gate signals are spin waves. (B) Output characteristics of the magnonic switch. The amplitude and the phase of the output spin wave depend on the amplitudes and phases of the source and gate spin waves. The Table below the output characteristics shows 5 selected output states to be used for multi-valued logic gate construction. (C) Truth Tables illustrating the logic operation of the magnonic interferometric switch.

Figure 3. (A) Symbols and truth tables of the two passive elements for independent phase and amplitude modulation: A-π phase shifter provides an a-π phase shift to the propagating spin wave, and an attenuator introduces an exponential damping to the propagating spin wave. (B) Schematics of the NOT and XOR 3VL gates built of the magnonic interferometric switch and the passive elements.

Figure 4. Photo of the YIG cross junction and connection schematics. The length of the cross is 3 mm; the width is 360 μm; and the YIG film thickness is 2.0 μm. There are four Π-shaped micro-antennas fabricated on the edges of the cross. Antennas are connected to a programmable network analyzer (PNA) Keysight N5241A. Two continuous spin wave signals are excited at terminals S and G. The inductive voltage is detected at terminal D. There are attenuators and a phase shifter aimed to equalize the amplitudes of the input waves and control the phase difference between the S and G terminals.

Figure 5. Experimental data demonstrating output voltage modulation by the phase difference between the S and G terminals. The operational frequency is 4.095 GHz, and bias magnetic field H = 798 Oe.  The red markers show experimentally measured output voltage. The red curve shows output voltage calculated in the ideal case of zero input phase/amplitude variation. The blue markers depict the phase of the output. The output phase is defined with respect to the phase of the spin wave generated at the terminal S. (A) Bias magnetic field is directed along the virtual S-D line. (B) Bias magnetic field is directed perpendicular to the virtual S-D line. (C) Bias magnetic field is directed at 45 degrees to the virtual S-D line. (D) Results for modified configuration, where the source and the gate antennas are located on the same arm of the cross as shown in the inset. The bias magnetic field $H$ is directed parallel to the virtual S-G line.

Figure 6. Experimental data for 15,000 subsequent measurements (150 points per 448 ms sweep) showing the variation of the output voltage in time at fixed input parameters. The markers depict the maximum output voltage measured for the On state (i.e. constructive interference). The blue markers show the variation of the output voltage for the Off state (destructive interference).

Figure 7. Results of micromagnetic simulations showing the propagation of spin waves in YIG cross at different operational frequency $f$. The direction of the bias magnetic field is the same as in Fig.5(B). The red arrow in each plot shows the direction of the input spin wave signal. (A) Data obtained for $f$ = 4.50 GHz. (B) Data obtained for $f$ = 4.64 GHz. (C) Data obtained for $f$ = 4.70 GHz.

Figure 8. Results of numerical modeling for nanometer-size switch with just 20 spins in the source-drain channel. The black curve depicts the amplitude of the output spin wave. The blue curve shows the phase of the amplitude.   (A) Numerical simulation at zero temperature. (B) Numerical simulations at room temperature.

Figure 9. Results of numerical modeling for the nanometer-size switch showing the noise strength as a function of the exchange coupling energy $J$.

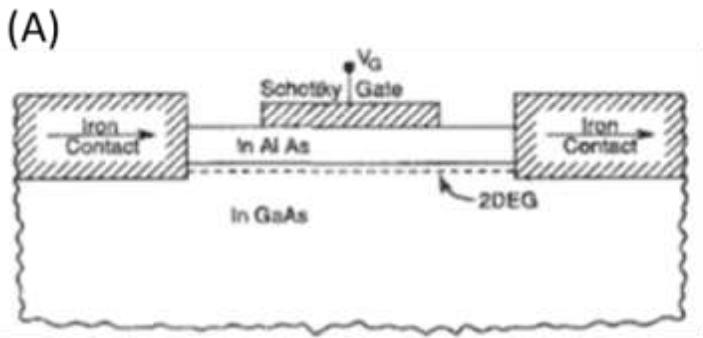
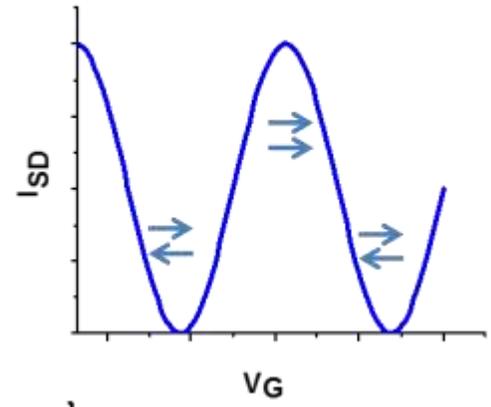
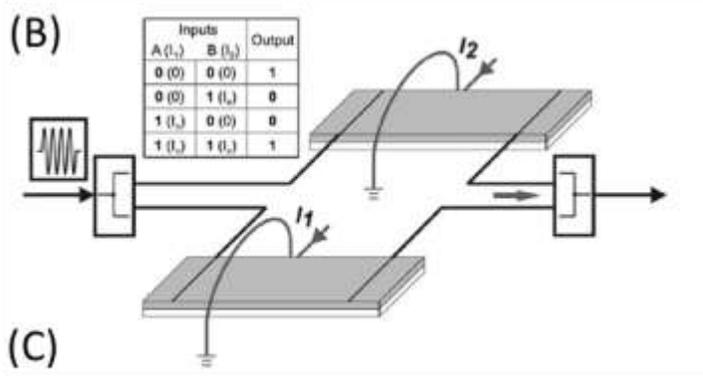
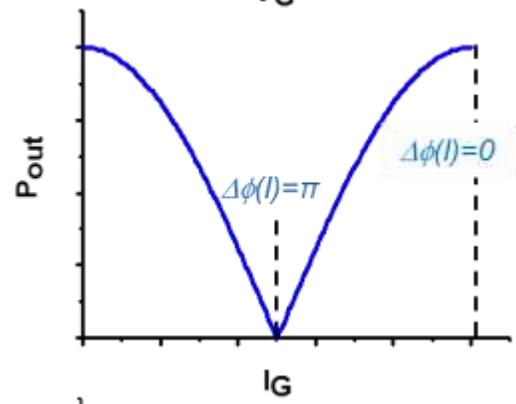
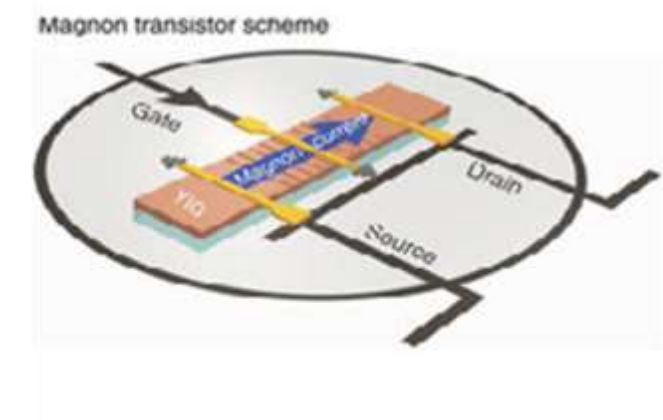
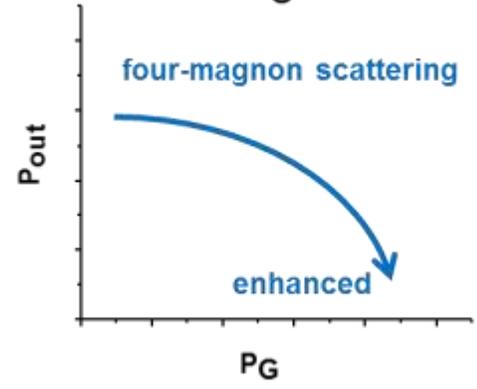

**Figure 1**

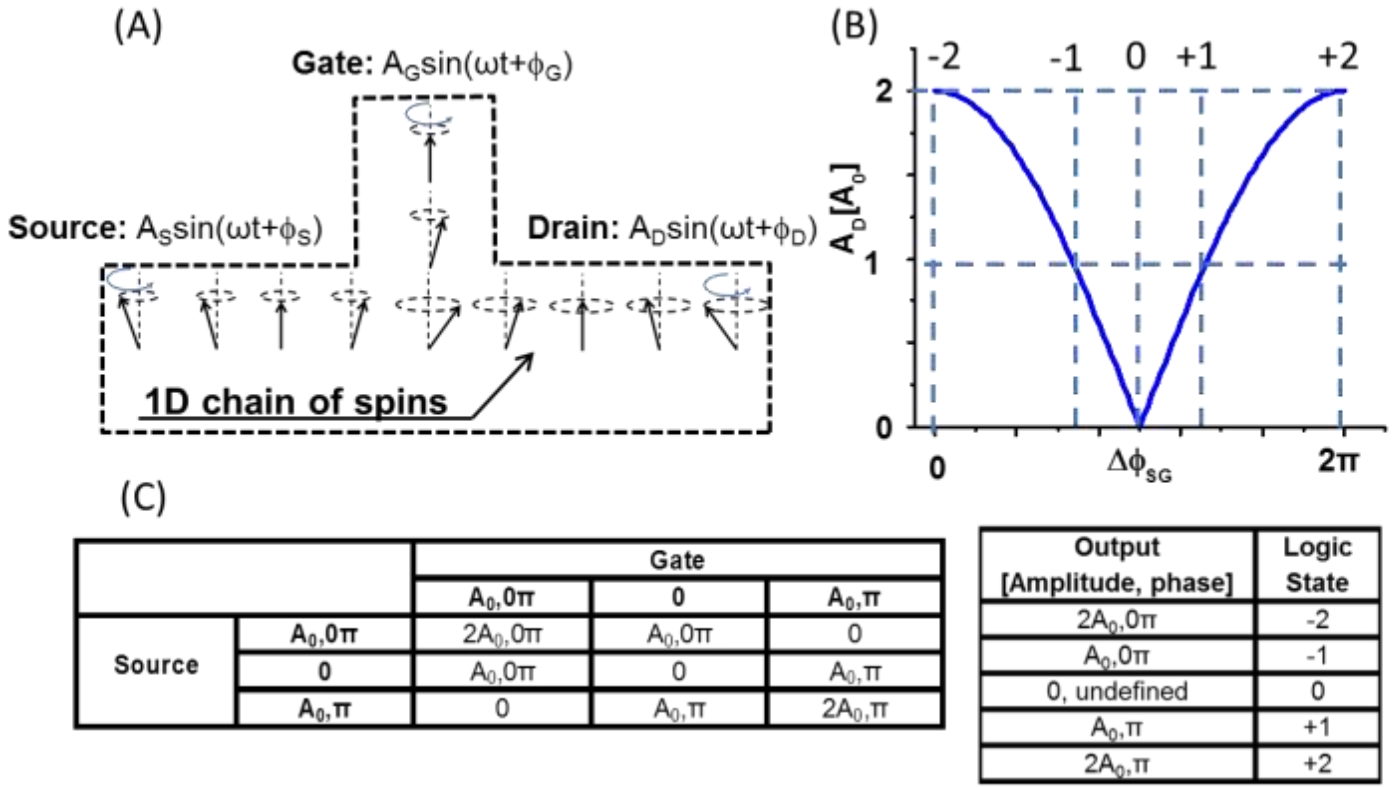

Figure 2

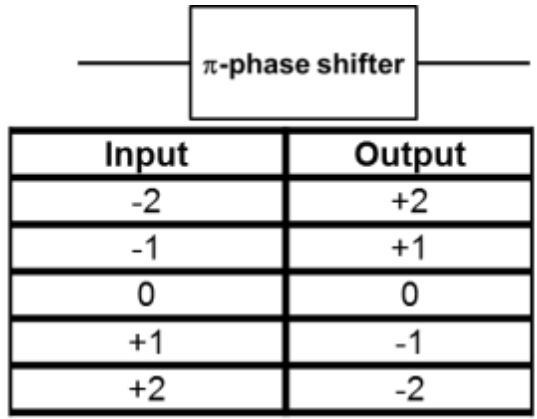
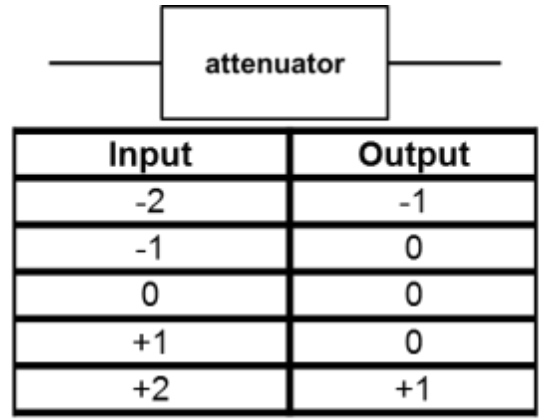
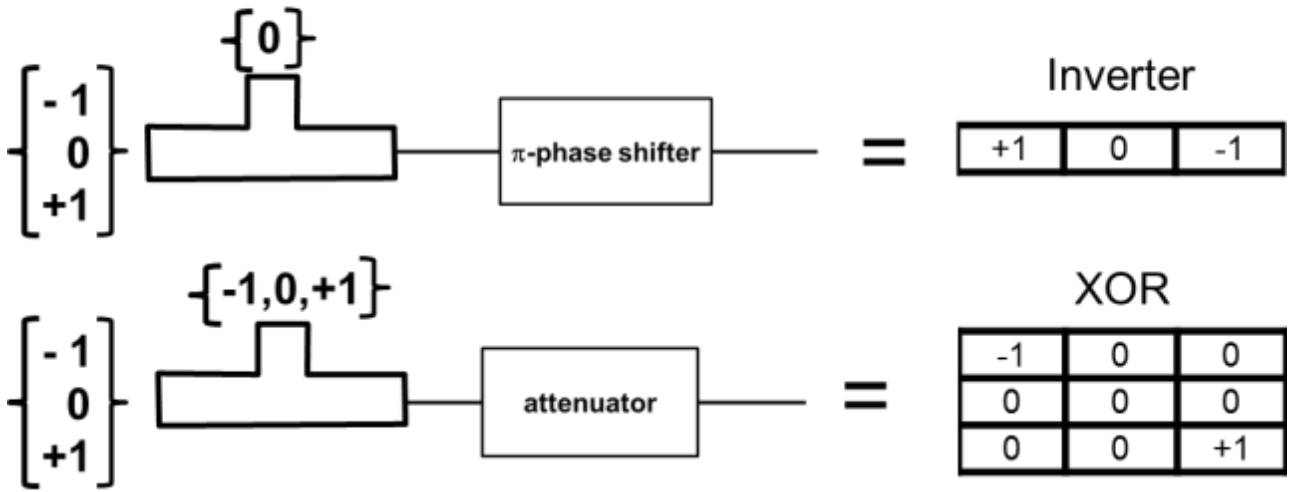

**Figure 3**

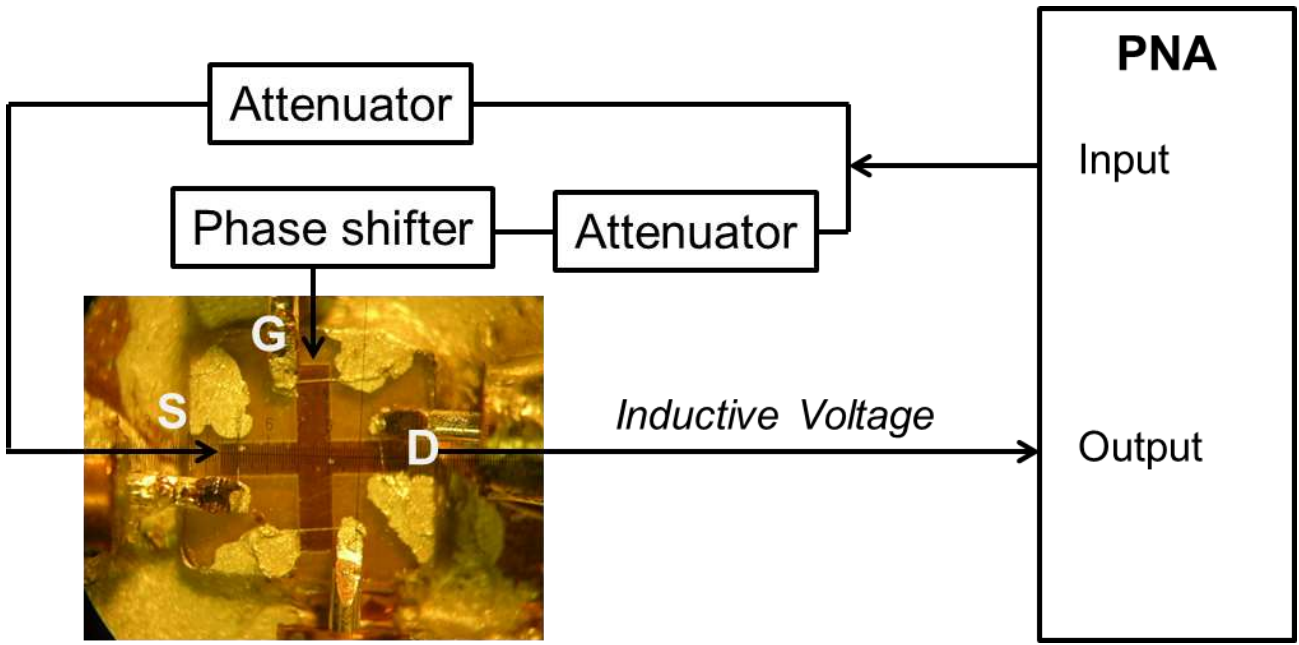

**Figure 4**

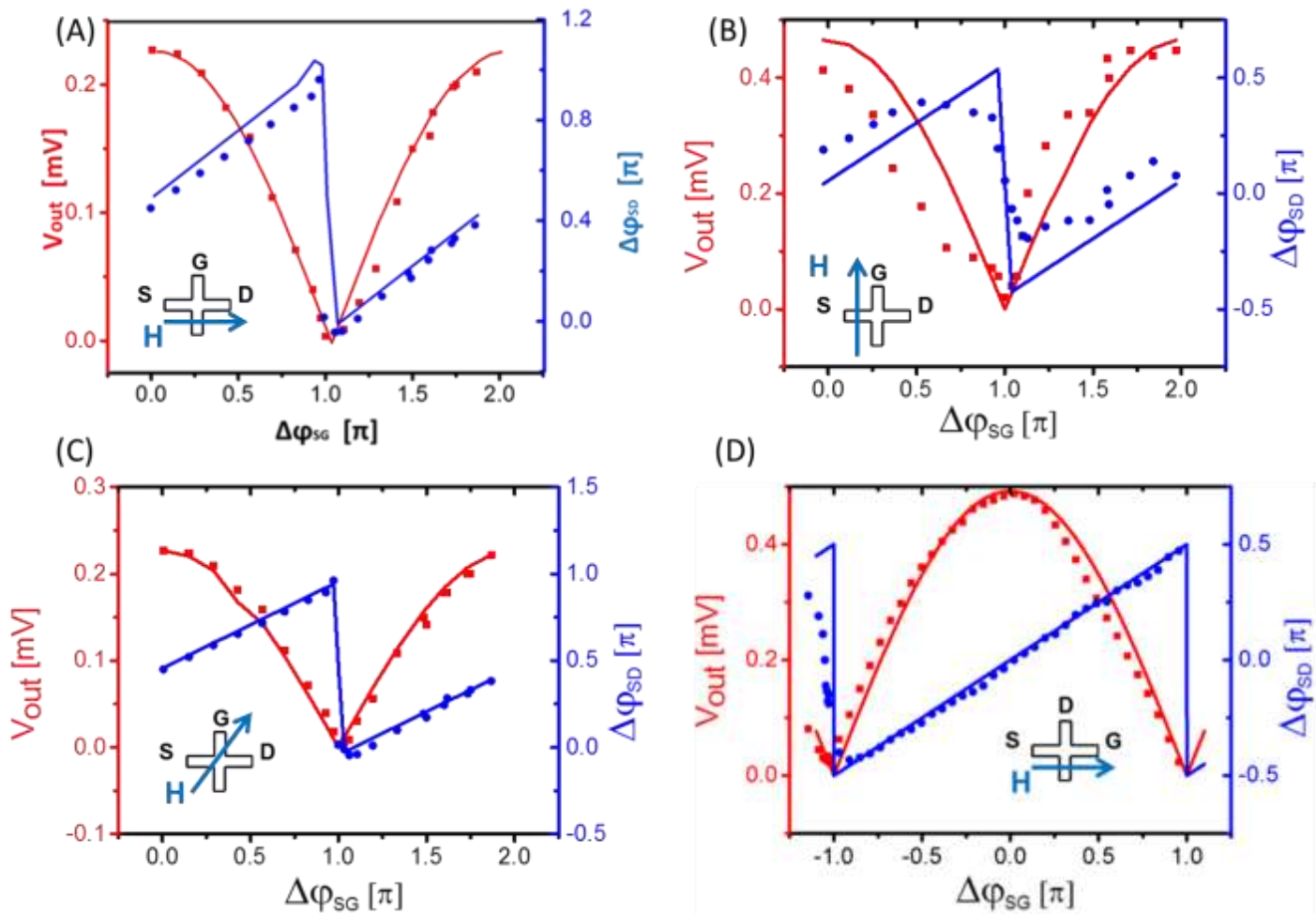

**Figure 5**

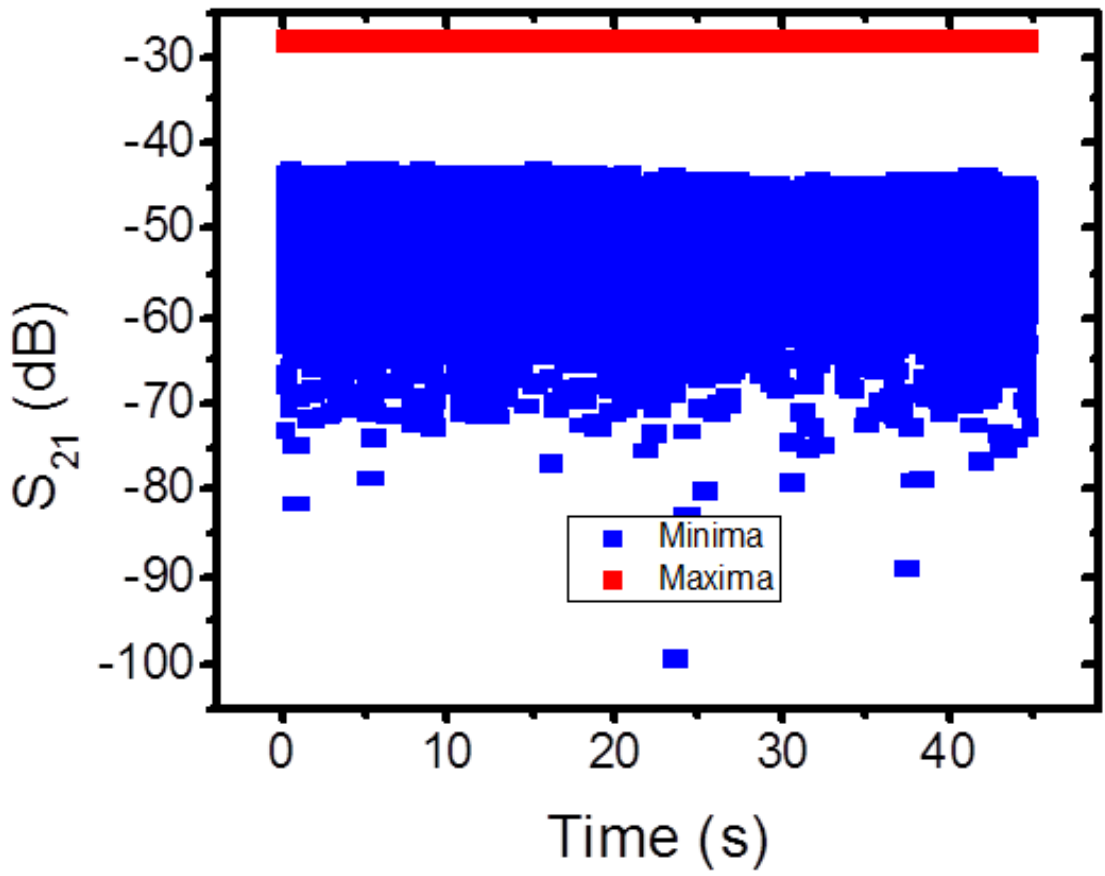

**Figure 6**

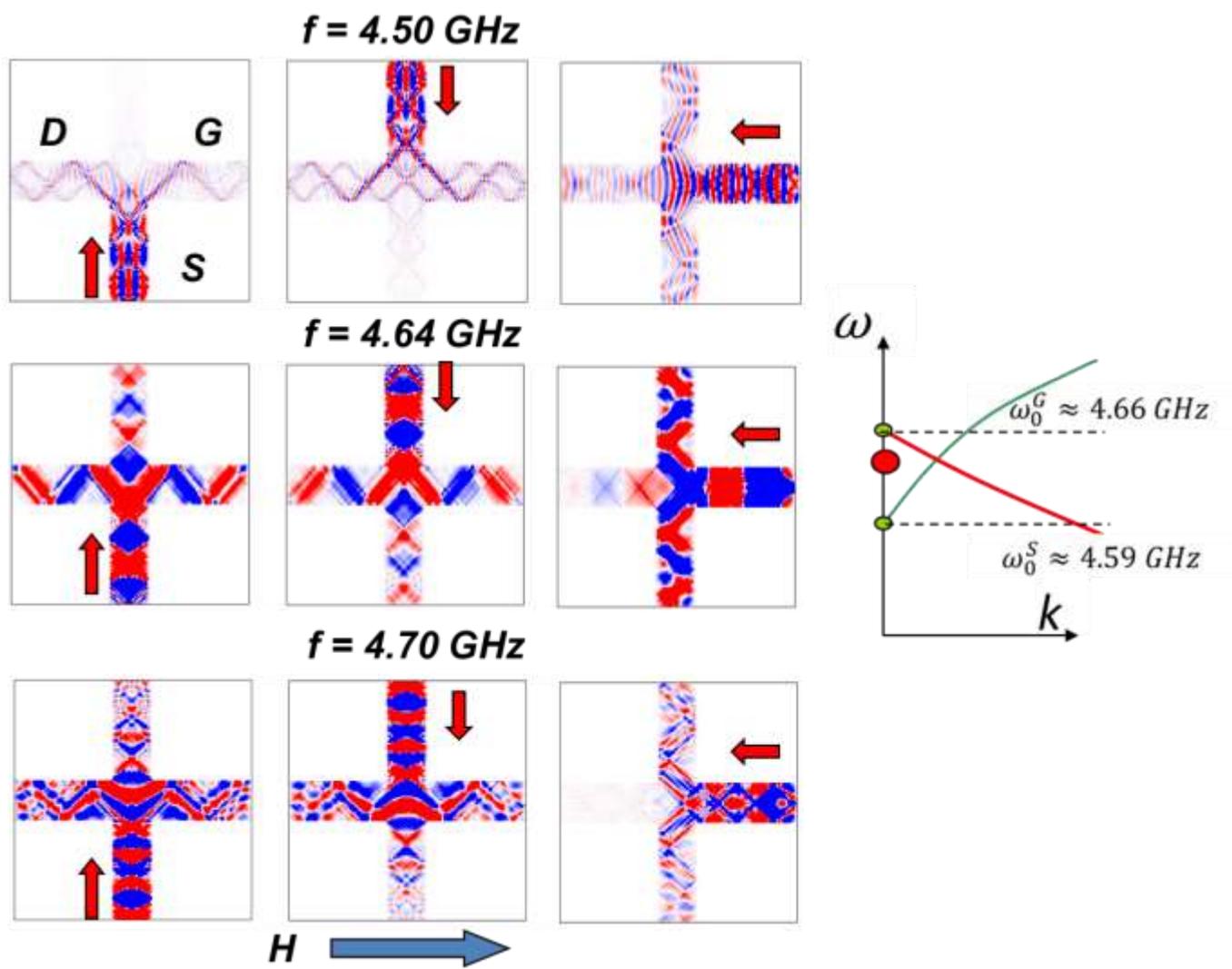

**Figure 7**

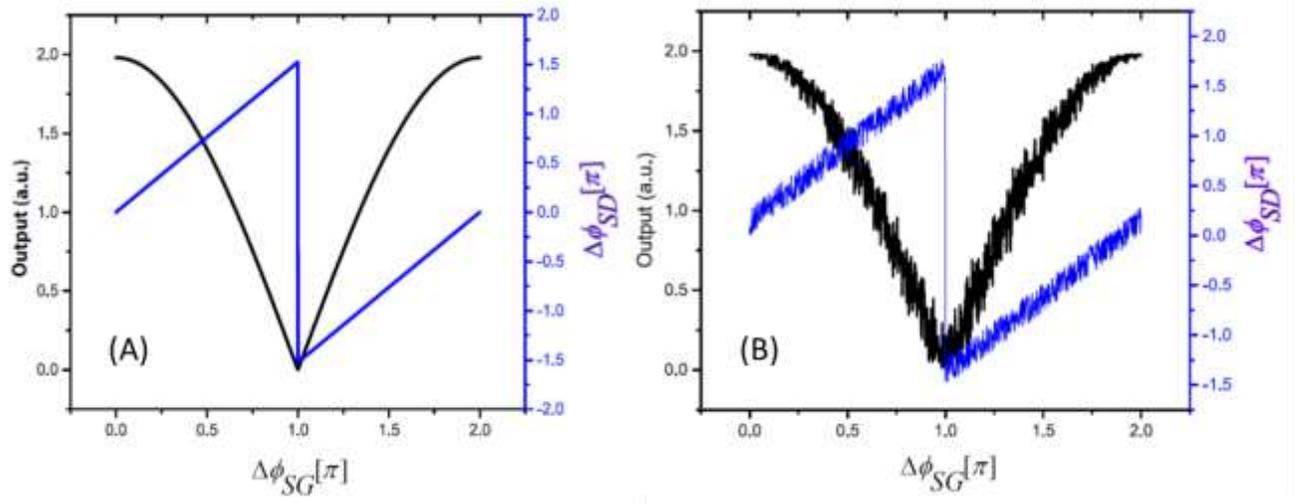

**Figure 8**

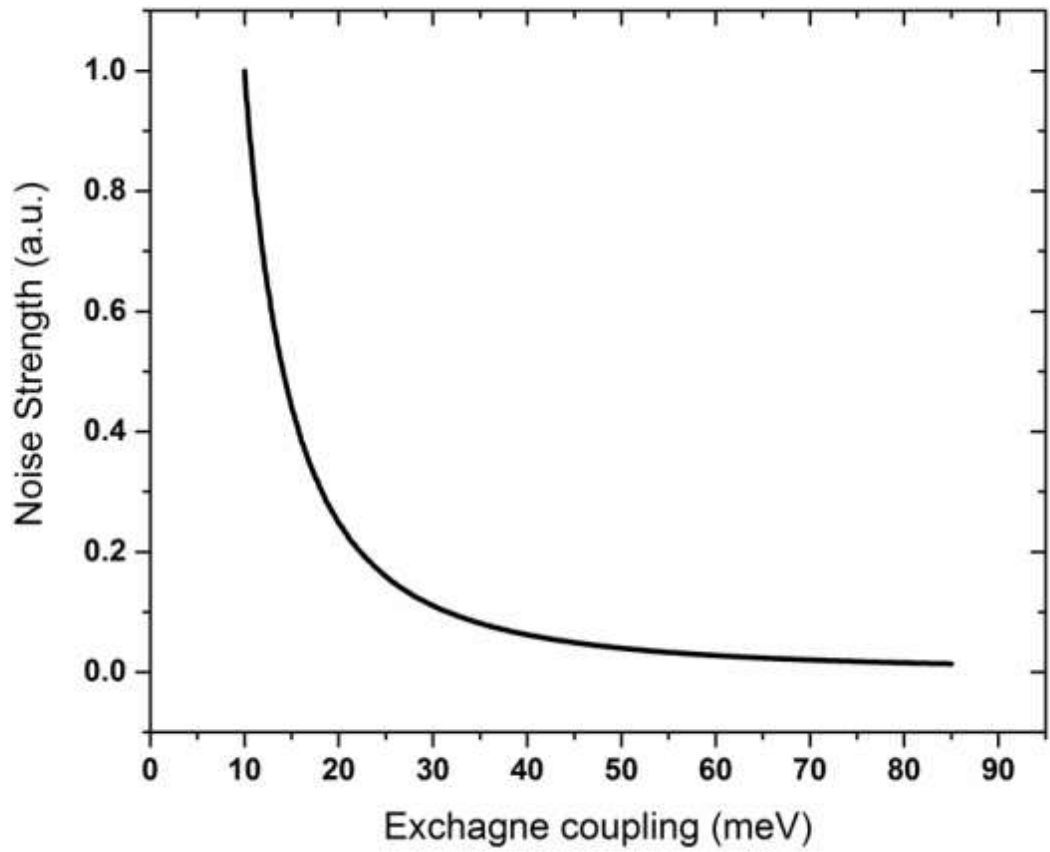

**Figure 9**